\documentclass[journal]{IEEEtran}

\usepackage[dvips]{color}
\usepackage{graphicx}
\usepackage{amsmath}

\begin{document}

\title{Physical-Layer Security for Spectrum Sharing Systems}

\markboth{IEEE Transactions on Wireless Communications (ACCEPTED TO APPEAR)}%
{Yulong Zou: Physical-Layer Security for Spectrum Sharing Systems}

\author{Yulong~Zou,~\IEEEmembership{Senior Member,~IEEE}

\thanks{Manuscript received April 8, 2016; revised September 13, 2016; accepted December 19, 2016. The editor coordinating the review of this paper and approving it for publication was Prof. X. Zhou.}

\thanks{Y. Zou is with the School of Telecommunications and Information Engineering, Nanjing University of Posts and Telecommunications, Nanjing 210003, Jiangsu, P. R. China. (Email: \{yulong.zou\}@njupt.edu.cn)}

\thanks{This work was partially supported by the National Natural Science Foundation of China (Grant Nos. 61401223 and 61522109), the Natural Science Foundation of Jiangsu Province (Grant Nos. BK20140887 and BK20150040), and the Key Project of Natural Science Research of Higher Education Institutions of Jiangsu Province (No. 15KJA510003).}

}

\maketitle

\vspace{-0.65cm}

\begin{abstract}
In this paper, we examine the physical-layer security for a spectrum sharing system consisting of {{multiple source-destination pairs, which dynamically access their shared spectrum for data transmissions in the presence of an eavesdropper.}} We propose a source cooperation (SC) aided opportunistic jamming framework for protecting the transmission confidentiality of the spectrum sharing system against eavesdropping. Specifically, when a source node is allowed to access the shared spectrum for data transmissions, another source is opportunistically selected in the spectrum sharing system to transmit an artificial noise for disrupting the eavesdropper without affecting the legitimate transmissions. We present two specific SC aided opportunistic jamming schemes, namely the SC aided random jammer selection (RJS) and optimal jammer selection (OJS), which are referred to as the SC-RJS and SC-OJS, respectively. We also consider the conventional non-cooperation as a baseline. We derive closed-form intercept probability expressions for the non-cooperation, SC-RJS and SC-OJS schemes, based on which their secrecy diversity gains are determined through an asymptotic intercept probability analysis in the high signal-to-noise ratio (SNR) region. It is proved that the conventional non-cooperation exhibits a secrecy diversity of zero, whereas the proposed SC-RJS and SC-OJS achieve a higher secrecy diversity of one. {{This also surprisingly means that no additional secrecy diversity gain is achieved by the optimal jammer selection compared to the random selection strategy.}} In addition, numerical results show that the intercept probability performance of the SC-OJS is always better than that of the SC-RJS and non-cooperation, even when the legitimate channel is worse than the eavesdropping channel.
\end{abstract}


\IEEEpeerreviewmaketitle

\section{Introduction}
Spectrum sharing allows heterogeneous wireless networks to coexist and access the same spectrum resource in a dynamic manner, also called dynamic spectrum access [1], [2], which has the advantage of increasing the spectrum utilization over the conventional static spectrum access. The concept of spectrum sharing was proposed in cognitive radio networks to enable an unlicensed wireless system to opportunistically access a licensed spectrum band, such as the TV band that is dedicated to broadcast television networks, but not used by the dedicated networks at a particular time, referred to as a TV white space [3]. As observed, the licensed television networks have higher priority than other unlicensed wireless networks in accessing their shared TV spectrum. Recently, spectrum sharing was examined for long term evolution (LTE) in unlicensed spectrum e.g. the 5GHz band which is populated by Wi-Fi devices [4], where different wireless networks should have the same priority for the spectrum access. Due to the broadcast nature of radio propagation, any active transmissions operated over the shared spectrum by different wireless networks may be readily overheard by an eavesdropper and is extremely vulnerable to eavesdropping [5]. It is therefore of importance to investigate the confidentiality protection of spectrum-sharing communications against eavesdropping attack.

Physical-layer security emerges as an effective means of securing wireless communications against eavesdropping by exploiting the physical characteristics of wireless channels [6]. It was proved in [7] that a source node can communicate with its destination in perfect secrecy from an information-theoretic perspective, when the main channel spanning from the source to destination has a better condition than the wiretap channel spanning from the source to eavesdropper. In [8], Leung-Yan-Cheong and Hellman introduced the notion of secrecy capacity which is shown as the difference between the capacity of the main channel and that of the wiretap channel. Later on, extensive research efforts were devoted to improving the secrecy capacity of wireless communications in fading environments by employing the artificial noise [9]-[12] and beamforming techniques [13]-[15]. More specifically, as discussed in [9]-[12], the artificial noise is a special signal designed in the null space of the main channel, which is emitted to interfere with the eavesdropper without affecting the legitimate destination. By contrast, beamforming techniques as studied in [13]-[15] enable the source to transmit its confidential signal in a particular direction to ensure that the received signals at the destination and eavesdropper experience constructive and destructive interference, respectively, thus leading to a significant performance improvement in terms of the secrecy capacity.

Recently, physical-layer security was further examined for cognitive radio networks [16], [17], where the rate of cognitive transmissions is maximized without causing any confidential information leakage to the eavesdropper. In [18] and [19], relay selection was studied for enhancing the physical-layer security of cognitive radio communications against eavesdropping. It was shown that the secrecy outage probability of cognitive transmission relying on relay selection is significantly reduced with an increasing number of relay nodes. In [20], multiuser scheduling was considered as an alternative means of improving the physical-layer security of cognitive transmissions and the corresponding secrecy capacity was evaluated over Rayleigh fading channels. More recently, in [21], we investigated the security-reliability tradeoff (SRT) for cognitive radio networks and proposed two relay selection schemes, namely the single-relay and multi-relay selection. Specifically, the single-relay selection chooses the ``best" relay only for assisting cognitive transmissions, whereas the multi-relay selection allows multiple relays to participate in protecting cognitive radio networks against eavesdropping.

In this paper, we explore physical-layer security for a spectrum sharing system, where {{multiple source-destination pairs}} share the same spectrum resource in the face of an eavesdropper. We consider that the eavesdropper constantly monitors the spectrum of interest and can overhear any confidential messages transmitted over the shared spectrum. The main contributions of this paper can be summarized as follows. First, we propose a source cooperation (SC) aided opportunistic jamming framework for improving the physical-layer security of spectrum sharing systems, where different source nodes cooperate with each other in defending against eavesdropping. Secondly, we present two specific SC aided opportunistic jamming schemes, namely the SC aided random jammer selection (RJS) and optimal jammer selection (OJS), denoted by the SC-RJS and SC-OJS, respectively. {{To be specific, when a source is scheduled to access the shared spectrum for transmitting to its destination, another source node is randomly chosen in the SC-RJS to emit an artificial noise for preventing the eavesdropper, whereas the SC-OJS would select the ``best" source node for protecting the transmission confidentiality against eavesdropping.}} Thirdly, we derive closed-form intercept probability expressions for the conventional non-cooperation as well as the proposed SC-RJS and SC-OJS schemes over Rayleigh fading channels. Finally, secrecy diversity gains of the non-cooperation, SC-RJS and SC-OJS schemes are characterized through an asymptotic intercept probability analysis in the high signal-to-noise ratio (SNR) region. We prove that the SC-RJS and SC-OJS schemes achieve a secrecy diversity of one, but the non-cooperation has a secrecy diversity of zero only, showing the secrecy benefit of proposed source cooperation framework in defending against eavesdropping.

The reminder of this paper is organized as follows. Section II presents the spectrum-sharing system model as well as proposes the SC-RJS and SC-OJS schemes. For comparison purposes, the conventional non-cooperation is also described in this section. Next, we derive closed-form intercept probability expressions for the non-cooperation, SC-RJS and SC-OJS schemes over Rayleigh fading channels in Section III, followed by Section IV, where the secrecy diversity analysis is presented. Then, numerical results are provided in Section V. Finally, Section VI gives some concluding remarks.

\section{Source Cooperation aided Opportunistic Jamming}
In this section, we first present the model of a general spectrum sharing system consisting of {{multiple source-destination pairs}}, which are allowed to dynamically share the same spectrum, while an eavesdropper is considered to be capable of overhearing and taping any active transmissions operated over the shared spectrum of interest. Then, a source cooperation (SC) aided opportunistic jamming framework is proposed for improving the physical-layer security of spectrum sharing system against eavesdropping.

\subsection{System Model and Problem Formulation}
\begin{figure}
  \centering
  {\includegraphics[scale=0.55]{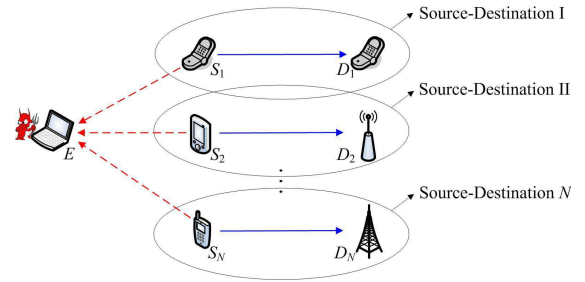}\\
  \caption{A general spectrum sharing system comprised of $N$ {{multiple source-destination pairs}} in the presence of a common eavesdropper (\emph{E}).}\label{Fig1}}
\end{figure}

As shown in Fig. 1, we consider a spectrum sharing system, where $N$ {{source-destination pairs coexist and dynamically share the same spectrum. Throughout this paper, we assume that the $N$ source-destination pairs are coordinated e.g. through a common spectrum database [23], [24], which guarantees that all the source nodes can orderly access their shared spectrum without signal interference. The design of a specific spectrum sharing policy [25] should consider both the spectrum efficiency and sharing fairness between different user pairs, which is beyond the scope of this paper. Although the focus of this paper is on the secrecy diversity analysis of coordinated source-destination pairs in the presence of an eavesdropper, similar secrecy diversity results can be obtained for the uncoordinated case, where different source-destination pairs may interfere with each other.}}

For notational convenience, let $H_i$ denote that the shared spectrum is allocated to the {{source-destination pair}} $i$, where $i$ is in the range from $1$ to $N$. To be specific, given $H_i$, it means that the {{source-destination pair}} $i$ is allowed to access the spectrum and the source $S_i$ starts to transmit to its destination $D_i$. Without loss of generality, let $\alpha_i = \Pr (H_i)$ represent the probability that the shared spectrum becomes available to the {{source-destination pair}} $i$, which can also be interpreted as the percentage of time period in which the {{source-destination pair}} $i$ is actively occupying over the shared spectrum, called \emph{duty cycle}. Clearly, the duty cycle $\alpha_i$ should be in the range from $0$ to $1$ and the sum of all the {{source-destination pairs}}' duty cycles should satisfy
\begin{equation}\label{equa1}
0 \le \sum\limits_{i = 1}^N {{\alpha _i}}  \le 1,
\end{equation}
where $N$ represents the number of {{source-destination pairs}}. Meanwhile, as shown in Fig. 1, an eavesdropper ($E$) is considered to tap any active transmissions operated over the spectrum shared by $N$ {{source-destination pairs}}. As a consequence, when the $S_i$ transmits to the $D_i$, the $E$ is assumed to be capable of overhearing the $S_i$-$D_i$ transmission. It is pointed out that all the wireless links between any two nodes of Fig. 1 are modeled as independent Rayleigh fading channels. In addition, any receiver of Fig. 1 is assumed to be deteriorated by a zero-mean additive white Gaussian noise (AWGN) with a variance of $N_0$.

Without loss of generality, we consider that the $S_i$ starts to transmit its signal $x_i$ at a power of $P_s$. Thus, the received signal at the $D_i$ can be written as
\begin{equation}\label{equa2}
y_i  = h_{s_id_i} \sqrt {P_s } x_i + n_i,
\end{equation}
where $h_{s_id_i}$ represents the fading gain of $S_i$-$D_i$ channel and $n_i$ represents the AWGN encountered at the $D_i$. Using the Shannon's capacity formula, we obtain an {{instantaneous capacity}} of $S_i$-$D_i$ transmission from (2) as
\begin{equation}\label{equa3}
C_{s_id_i}  = \log _2 ( {1 + {{|h_{s_id_i} |^2 \gamma_s }}} ),
\end{equation}
where $\gamma_s  = {{P_s }}/{{N_0 }}$ is referred to as the signal-to-noise ratio (SNR). Meanwhile, due to the broadcast nature of radio propagation, the $E$ also overhears the signal transmission of $S_i$ and thus the corresponding received signal is expressed as
\begin{equation}\label{equa4}
y_e  = h_{s_ie} \sqrt {P_s } x_i + n_e,
\end{equation}
where $h_{s_ie}$ represents the fading gain of $S_i$-$E$ channel and $n_e$ represents the AWGN encountered at the $E$. Similarly to (3), an {{instantaneous capacity}} of the wiretap channel from the $S_i$ to $E$ is given by
\begin{equation}\label{equa5}
C_{s_ie}  = \log _2 ( {1 + {{|h_{s_ie} |^2 \gamma_s }}} ).
\end{equation}

Following the physical-layer security literature [8]-[16], a perfect secrecy can be achieved only when an {{instantaneous capacity}} of the main channel $C_{s_id_i}$ (spanning from $S_i$ to $D_i$) is higher than that of the wiretap channel $C_{s_ie}$ (spanning from $S_i$ to $E$). If an {{instantaneous capacity}} of the main channel $C_{s_id_i}$ drops below that of the wiretap channel $C_{s_ie}$, the $E$ would be capable of successfully decoding the source signal and an intercept event is considered to happen [19]. In this paper, the probability of occurrence of an intercept event (referred to as intercept probability) is used to measure the physical-layer security of spectrum sharing systems.

\subsection{SC aided Opportunistic Jamming}
In this section, we propose the use of so-called SC aided opportunistic jamming for protecting the spectrum sharing system against eavesdropping, where the $N$ {{source-destination pairs}} of Fig. 1 are enabled to cooperate with each other. To be specific, when a {{source node}} is allowed to access the spectrum for data transmissions, another {{source}} may be opportunistically selected to act as a friendly jammer for interfering with the $E$ without affecting the legitimate transmissions. For notational convenience, let ${\cal S}=\{S_1,S_2,\cdots,S_N\}$ denote the set of $N$ source nodes of the spectrum sharing system, as shown in Fig. 1. Without loss of generality, we consider that the $S_i$ is scheduled to access the spectrum and starts to transmit its signal $x_i$. In order to protect the source transmission, a friendly jammer denoted by $J$ is opportunistically chosen among the remaining idle source nodes to emit an artificial noise for confusing the $E$. Note that the total transmit power of the source $S_i$ and the selected friendly jammer $S_j$ is constrained to $P_s$. For simplicity, we consider the equal power allocation here and thus the transmit powers of the $S_i$ and $S_j$ are given by $P_s/2$.

{{In this paper, we assume that the artificial noise transmitted by the selected friendly jammer is generated from a pseudo random sequence, which is known to the legitimate receiver and remains unknown to the eavesdropper. Thus, the legitimate receiver $D_i$ is able to cancel out the artificial noise, while the $E$ is severely interfered. It is worth mentioning that the objective of this paper is to reveal the impact of jammer selection on the secrecy diversity of wireless communications and the artificial noise design is not our focus.}} Therefore, we can express the received signal at $D_i$ as
\begin{equation}\label{equa6}
y_i  = h_{s_id_i} \sqrt {\frac{P_s}{2} } x_i + n_i,
\end{equation}
from which an {{instantaneous capacity}} of $S_i$-$D_i$ transmission relying on the SC aided opportunistic jamming is obtained as
\begin{equation}\label{equa7}
C^{\textrm{SC}}_{s_id_i}  =  \log _2 ( {1 + {{|h_{s_id_i} |^2 \frac{\gamma_s}{2} }}} ).
\end{equation}
\quad Meanwhile, due to the broadcast nature of radio propagation, the $E$ can also overhear the $S_i$'s transmission. In order to defend against eavesdropping, another source node denoted by $S_j$ may be selected to act as a friendly jammer, which is employed to emit an artificial noise denoted by $x_n$ at a power of $P_s/2$ for confusing the $E$. {{Again, the artificial noise $x_n$ is pre-shared and known to the legitimate receiver so that the $D_i$ can cancel out $x_n$, as implied from (6). By contrast, the artificial noise $x_n$ is assumed to be unknown to the eavesdropper which would be interfered.}} Hence, the received signal at the $E$ can be expressed as
\begin{equation}\label{equa8}
y_e  = h_{s_ie} \sqrt {\frac{P_s}{2} } x_i +  h_{s_je} \sqrt {\frac{P_s}{2} }  x_n + n_e,
\end{equation}
where $h_{s_ie}$ and $h_{s_je}$ represent the fading gains of the channel from $S_i$ to $E$ and that from $S_j$ to $E$, respectively.
Using (8), we obtain an {{instantaneous capacity}} of the wiretap channel from the $S_i$ to $E$ with the aid of the selected friendly jammer $S_j$ as
\begin{equation}\label{equa9}
C^{\textrm{SC}}_{s_ie}(s_j)  = \log _2 ( {1 + \frac{|h_{s_ie} |^2\gamma_s}{|h_{s_je} |^2\gamma_s+2}}),
\end{equation}
where $S_j \in \{{\cal S}-S_i\}$ and $\{{\cal S}-S_i\}$ denotes the set of source nodes ${\cal S}$ excluding a set element $S_i$. In this paper, we consider two opportunistic jammer selection strategies, namely the random jammer selection (RJS) and optimal jammer selection (OJS). To be specific, in the RJS scheme, a source node in the set $\{{\cal S}-S_i\}$ is randomly chosen to act as the friendly jammer, whereas the OJS aims to minimize the confidential information leakage as much as possible. Hence, the RJS criterion is described as
\begin{equation}\label{equa10}
J = \mathop {{\rm{rand}}}\limits_{{S_j} \in \{ {\cal S} - {S_i}\} } {S_j},
\end{equation}
where ${\rm{rand}}(\cdot)$ denotes the equiprobable selection of an element from the set $\{ {\cal S} - {S_i}\}$. By contrast, in the OJS scheme, a source node $S_j$ that minimizes an {{instantaneous capacity}} of the wiretap channel $C^{\textrm{SC}}_{s_ie}(s_j) $ is used to act as the friendly jammer. By using (9), the OJS criterion can thus be written as
\begin{equation}\label{equa11}
J = \mathop {\min }\limits_{{S_j} \in \{ {\cal S} - {S_i}\} } C_{{s_i}e}^{{\textrm{SC}}}({s_j}) = \mathop {\max }\limits_{{S_j} \in \{ {\cal S} - {S_i}\} } |{h_{{s_j}e}}{|^2}.
\end{equation}
Combining (9) and (10), we obtain an {{instantaneous capacity}} of the wiretap channel from $S_i$ to $E$ with the aid of the RJS as
\begin{equation}\label{equa12}
C^{\textrm{SC-RJ}}_{s_ie}  = \mathop {\rm{rand} }\limits_{{S_j} \in \{ {\cal S} - {S_i}\} } \log _2 ( {1 + \frac{|h_{s_ie} |^2\gamma_s}{|h_{s_je} |^2\gamma_s+2}}),
\end{equation}
{{where the eavesdropper's channel state information (CSI) $h_{s_je}$ is not needed in performing the random jammer selection.}} Similarly, an {{instantaneous capacity}} of the $S_i$-$E$ channel with the help of the optimal jammer can be obtained from (9) and (11) as
\begin{equation}\label{equa13}
C^{\textrm{SC-OJ}}_{s_ie}  = \mathop {\min }\limits_{{S_j} \in \{ {\cal S} - {S_i}\} } \log _2 ( {1 + \frac{|h_{s_ie} |^2\gamma_s}{|h_{s_je} |^2\gamma_s+2}}),
\end{equation}
which shows that the eavesdropper's CSI $h_{s_je}$ is required to carry out the optimal jammer selection for the sake of minimizing the confidential information leakage. Since all the wireless links between any two nodes of Fig. 1 are modeled as independent Rayleigh fading channels, the random variables of $|h_{s_id_i} |^2$, $|h_{s_ie} |^2$ and $|h_{s_je} |^2$ are exponentially distributed with respective means of $\sigma^2_{s_id_i}$, $\sigma^2_{s_ie}$ and $\sigma^2_{s_je}$, respectively. It is pointed out that the average fading gains $\sigma^2_{s_id_i}$, $\sigma^2_{s_ie}$ and $\sigma^2_{s_je}$ may be different due to the fact that the sources, destinations and eavesdropper move around and experience different path losses.

\section{Intercept Probability Analysis over Rayleigh Fading Channels}
In this section, we analyze the intercept probability of SC-RJS and SC-OJS schemes over Rayleigh fading channels. For comparison purposes, we also conduct the intercept probability analysis of conventional non-cooperation for spectrum sharing systems.

\subsection{Conventional Non-cooperation}
In conventional non-cooperation scheme, when the shared spectrum is assigned to a {{source-destination pair}} $i$, the $S_i$ starts to transmit its confidential information to its destination $D_i$. As aforementioned, an intercept event is considered to occur when an {{instantaneous capacity}} of the main channel $C_{s_id_i}$ falls below that of the wiretap channel $C_{s_ie}$. Note that there are $N$ {{source-destination pairs}} orderly accessing their shared spectrum. Hence, using the law of total probability, we obtain an intercept probability of the spectrum sharing system relying on the non-cooperation scheme as
\begin{equation}\label{equa14}
\begin{split}
P_{{\mathop{\textrm{int}}} }^{{\textrm{nonC}}} &= \sum\limits_{i = 1}^N {\Pr ({H_i})\Pr ({C_{{s_i}{d_i}}} < {C_{{s_i}e}})}  \\
&= \sum\limits_{i = 1}^N {{\alpha _i}\Pr ({C_{{s_i}{d_i}}} < {C_{{s_i}e}})},
\end{split}
\end{equation}
where $N$ is the number of {{source-destination pairs}} and $\alpha_i$ denotes the duty cycle of the {{source-destination pair}} $i$. Substituting ${C_{{s_i}{d_i}}} $ and ${C_{{s_i}{e}}}$ from (3) and (5) into (14) gives
\begin{equation}\label{equa15}
P_{{\mathop{\textrm{int}}} }^{{\textrm{nonC}}} = \sum\limits_{i = 1}^N {{\alpha _i}\Pr (|{h_{{s_i}{d_i}}}{|^2} < |{h_{{s_i}e}}{|^2})}.
\end{equation}
Noting that fading gains $|{h_{{s_i}{d_i}}}|$ and $|{h_{{s_i}{e}}}|$ are modeled as Rayleigh random variables, we can obtain that $|{h_{{s_i}{d_i}}}|^2$ and $|{h_{{s_i}{e}}}|^2$ are exponentially distributed. Letting ${\sigma}^2_{{s_id_i}}$ and ${{\sigma}^2_{{s_ie}}}$ denote the means of $|{h_{{s_i}{d_i}}}|^2$ and $|{h_{{s_i}{e}}}|^2$, respectively, we have
\begin{equation}\label{equa16}
P_{{\mathop{\textrm{int}}} }^{{\textrm{nonC}}} = \sum\limits_{i = 1}^N {\frac{{{\alpha _i}\sigma _{{s_i}e}^2}}{{\sigma _{{s_i}{d_i}}^2 + \sigma _{{s_i}e}^2}}}, \end{equation}
which gives a closed-form intercept probability of the conventional non-cooperation scheme for spectrum sharing systems in the presence of an eavesdropper. It can be observed from (16) that the intercept probability only relates to the duty cycle $\alpha_i$ as well as the average channel gains $\sigma^2_{s_id_i}$ and $\sigma^2_{s_ie}$, but is independent of the SNR $\gamma_s$.

\subsection{SC-RJS Scheme}
This subsection presents the intercept probability analysis of SC-RJS scheme. Similarly to (14), an intercept probability of the SC-RJS scheme is obtained as
\begin{equation}\label{equa17}
P_{{\mathop{\textrm{int}}} }^{{\textrm{SC-RJ}}} = \sum\limits_{i = 1}^N {{\alpha _i}\Pr (C_{{s_i}{d_i}}^{{\textrm{SC}}} < C_{{s_i}e}^{{\textrm{SC-RJ}}})},
\end{equation}
where $C_{{s_i}{d_i}}^{{\textrm{SC}}} $ and $C_{{s_i}e}^{{\textrm{SC-RJ}}}$ are given by (7) and (12), respectively. Combining (7), (12) and (17), we arrive at
\begin{equation}\label{equa18}
\begin{split}
P_{{\mathop{\textrm{int}}} }^{{\textrm{SC-RJ}}} =& \sum\limits_{i = 1}^N {{\alpha _i}\sum\limits_{{S_j} \in \{ {\cal S} - {S_i}\} } {\Pr (\frac{{|{h_{{s_i}{d_i}}}{|^2}{\gamma _s}}}{2} < \frac{{|{h_{{s_i}e}}{|^2}{\gamma _s}}}{{|{h_{{s_j}e}}{|^2}{\gamma _s} + 2}}} }\\
&\quad \quad \quad \quad \quad \quad \quad \quad,J = {S_j}).
\end{split}
\end{equation}
As observed from (10), in the RJS, each source node in the set $\{ {\cal S} - {S_i}\}$ has an equal chance to be selected as the friendly jammer. Moreover, the RJS process is independent of random variables $|{h_{{s_i}{d_i}}}{|^2}$, $|{h_{{s_i}{e}}}{|^2}$, and $|{h_{{s_j}{e}}}{|^2}$. Therefore, we can simplify (18) as
\begin{equation}\label{equa19}
\begin{split}
P_{{\mathop{\textrm{int}}} }^{{\textrm{SC-RJ}}} = &\sum\limits_{i = 1}^N {\frac{{{\alpha _i}}}{{N - 1}}\sum\limits_{{S_j} \in \{ {\cal S} - {S_i}\} } {\Pr (\frac{{|{h_{{s_i}{d_i}}}{|^2}{\gamma _s}}}{2} } }\\
&\quad \quad \quad \quad \quad \quad< \frac{{|{h_{{s_i}e}}{|^2}{\gamma _s}}}{{|{h_{{s_j}e}}{|^2}{\gamma _s} + 2}}),
\end{split}
\end{equation}
which is given by
\begin{equation}\label{equa20}
P_{{\mathop{\textrm{int}}} }^{{\textrm{SC-RJ}}} = \sum\limits_{i = 1}^N {\frac{{{\alpha _i}}}{{N - 1}}\sum\limits_{{S_j} \in \{ {\cal S} - {S_i}\} } {\Pr (|{h_{{s_j}e}}{|^2}{\gamma _s} + 2 < \frac{{2|{h_{{s_i}e}}{|^2}}}{{|{h_{{s_i}{d_i}}}{|^2}}})} }.
\end{equation}
Denoting $|{h_{{s_i}e}}{|^2} = X$, $|{h_{{s_i}{d_i}}}{|^2} = Y$, and $Z = \frac{X}{Y}$, we can rewrite (20) as
\begin{equation}\label{equa21}
P_{{\mathop{\textrm{int}}} }^{{\textrm{SC-RJ}}} = \sum\limits_{i = 1}^N {\frac{{{\alpha _i}}}{{N - 1}}\sum\limits_{{S_j} \in \{ {\cal S} - {S_i}\} } {\Pr (|{h_{{s_j}e}}{|^2}{\gamma _s} + 2 < 2Z)} }.
\end{equation}
Meanwhile, the cumulative distribution function (CDF) of random variable $Z$ is expressed as
\begin{equation}\label{equa22}
\Pr (Z < z) = \Pr (X < zY),
\end{equation}
for $z>0$. Noting that $X$ and $Y$ are independent and exponentially distributed, we obtain the CDF of $Z$ as
\begin{equation}\label{equa23}
\begin{split}
\Pr (Z < z) = &\int_0^\infty  {\frac{1}{{\sigma _{{s_i}e}^2}}\exp ( - \frac{x}{{\sigma _{{s_i}e}^2}})dx}\\
&\int_{\frac{x}{z}}^\infty  {\frac{1}{{\sigma _{{s_i}{d_i}}^2}}\exp ( - \frac{y}{{\sigma _{{s_i}{d_i}}^2}})dy} ,
\end{split}
\end{equation}
where ${\sigma _{{s_i}{d_i}}^2}$ and ${\sigma _{{s_i}{e}}^2}$ are the respective means of $|{h_{{s_i}{d_i}}}{|^2}$ and $|{h_{{s_i}{e}}}{|^2}$. Using (23), we have
\begin{equation}\label{equa24}
\begin{split}
\Pr (Z < z)& = \int_0^\infty  {\frac{1}{{\sigma _{{s_i}e}^2}}\exp ( - \frac{x}{{\sigma _{{s_i}e}^2}} - \frac{x}{{\sigma _{{s_i}{d_i}}^2z}})dx}  \\
&= \frac{{\sigma _{{s_i}{d_i}}^2z}}{{\sigma _{{s_i}{d_i}}^2z + \sigma _{{s_i}e}^2}},
\end{split}
\end{equation}
from which the probability density function (PDF) of $Z$ is given by
\begin{equation}\label{equa25}
{p_Z}(z) = \frac{{\sigma _{{s_i}{d_i}}^2\sigma _{{s_i}e}^2}}{{{{(\sigma _{{s_i}{d_i}}^2z + \sigma _{{s_i}e}^2)}^2}}},
\end{equation}
for $z > 0$. Note that $|{h_{{s_j}e}}|$ is Rayleigh distributed, implying that $|{h_{{s_j}e}}{|^2}$ follows exponential distribution with a mean of ${\sigma^2_{{s_j}e}}$. Since $|{h_{{s_j}e}}{|^2}$ is independent of random variable $Z$, we can obtain the term $\Pr (|{h_{{s_j}e}}{|^2}{\gamma _s} + 2 < 2Z)$ as (26) at the top of the following page,
\begin{figure*}
\begin{equation}\label{equa26}
\begin{split}
 \Pr (|{h_{{s_j}e}}{|^2}{\gamma _s} + 2 < 2Z) &= \int_1^\infty  {\frac{{\sigma _{{s_i}{d_i}}^2\sigma _{{s_i}e}^2}}{{{{(\sigma _{{s_i}{d_i}}^2z + \sigma _{{s_i}e}^2)}^2}}}[1 - \exp ( - \frac{{2z - 2}}{{\sigma _{{s_j}e}^2{\gamma _s}}})]dz}  \\
&= \int_1^\infty  {\frac{{\sigma _{{s_i}{d_i}}^2\sigma _{{s_i}e}^2}}{{{{(\sigma _{{s_i}{d_i}}^2z + \sigma _{{s_i}e}^2)}^2}}}dz}  - \int_1^\infty  {\frac{{\sigma _{{s_i}{d_i}}^2\sigma _{{s_i}e}^2}}{{{{(\sigma _{{s_i}{d_i}}^2z + \sigma _{{s_i}e}^2)}^2}}}\exp ( - \frac{{2z - 2}}{{\sigma _{{s_j}e}^2{\gamma _s}}})dz}  \\
&= \frac{{\sigma _{{s_i}e}^2}}{{\sigma _{{s_i}{d_i}}^2 + \sigma _{{s_i}e}^2}} - \Omega (\sigma _{{s_i}{d_i}}^2,\sigma _{{s_i}e}^2,{\gamma _s}), \\
\end{split}
\end{equation}
\end{figure*}
where the parameter $\Omega (\sigma _{{s_i}{d_i}}^2,\sigma _{{s_i}e}^2,{\gamma _s})$ is given by
\begin{equation}\label{equa27}
\Omega (\sigma _{{s_i}{d_i}}^2,\sigma _{{s_i}e}^2,{\gamma _s}) = \int_1^\infty  {\frac{{\sigma _{{s_i}{d_i}}^2\sigma _{{s_i}e}^2}}{{{{(\sigma _{{s_i}{d_i}}^2z + \sigma _{{s_i}e}^2)}^2}}}\exp ( - \frac{{2z - 2}}{{\sigma _{{s_j}e}^2{\gamma _s}}})dz}.
\end{equation}
Denoting $\frac{{2z}}{{\sigma _{{s_j}e}^2{\gamma _s}}} + \frac{{2\sigma _{{s_i}e}^2}}{{\sigma _{{s_i}{d_i}}^2\sigma _{{s_j}e}^2{\gamma _s}}} =   t$, we can obtain $\Omega (\sigma _{{s_i}{d_i}}^2,\sigma _{{s_i}e}^2,{\gamma _s})$ as
\begin{equation}\label{equa28}
\begin{split}
&\Omega (\sigma _{{s_i}{d_i}}^2,\sigma _{{s_i}e}^2,{\gamma _s}) = \exp (\frac{{2\sigma _{{s_i}e}^2 + 2\sigma _{{s_i}{d_i}}^2}}{{\sigma _{{s_i}{d_i}}^2\sigma _{{s_j}e}^2{\gamma _s}}})\\
&\quad\times\int_{ \frac{2}{{\sigma _{{s_j}e}^2{\gamma _s}}} + \frac{{2\sigma _{{s_i}e}^2}}{{\sigma _{{s_i}{d_i}}^2\sigma _{{s_j}e}^2{\gamma _s}}}}^{ \infty } {\frac{{\sigma _{{s_i}e}^2}}{{\sigma _{{s_i}{d_i}}^2\sigma _{{s_j}e}^2{\gamma _s}{t^2}}}\exp (-t)dt},
\end{split}
\end{equation}
which is rewritten as
\begin{equation}\label{equa29}
\Omega (\sigma _{{s_i}{d_i}}^2,\sigma _{{s_i}e}^2,{\gamma _s}) = \exp (\varphi )\int_{ \varphi }^{ \infty } {\frac{{2\sigma _{{s_i}e}^2}}{{\sigma _{{s_i}{d_i}}^2\sigma _{{s_j}e}^2{\gamma _s}{t^2}}}\exp (-t)dt},
\end{equation}
where the parameter $\varphi$ is defined as
\begin{equation}\label{equa30}
\varphi  = \frac{2}{{\sigma _{{s_j}e}^2{\gamma _s}}} + \frac{{2\sigma _{{s_i}e}^2}}{{\sigma _{{s_i}{d_i}}^2\sigma _{{s_j}e}^2{\gamma _s}}}.
\end{equation}
Performing the partial integration to (29), we arrive at
\begin{equation}\label{equa31}
\begin{split}
& \Omega (\sigma _{{s_i}{d_i}}^2,\sigma _{{s_i}e}^2,{\gamma _s}) = \exp (\varphi )\int_{ \varphi }^{ \infty } {\frac{{2\sigma _{{s_i}e}^2}}{{\sigma _{{s_i}{d_i}}^2\sigma _{{s_j}e}^2{\gamma _s}}}\exp (-t)d( - {t^{ - 1}})}  \\
&= \frac{{2\sigma _{{s_i}e}^2}}{{\sigma _{{s_i}{d_i}}^2\sigma _{{s_j}e}^2{\gamma _s}}}\exp (\varphi )[\frac{1}{\varphi }\exp ( - \varphi ) - \int_{ \varphi }^{ \infty }{\frac{{\exp (-t)}}{t}dt} ] \\
&= \frac{{2\sigma _{{s_i}e}^2}}{{\sigma _{{s_i}{d_i}}^2\sigma _{{s_j}e}^2{\gamma _s}}}[\frac{1}{\varphi } - \exp (\varphi )Ei( \varphi )],
 \end{split}
\end{equation}
where $Ei(\varphi ) = \int_\varphi^{ \infty }{\frac{{{e^{-t}}}}{t}dt}  $ is known as the exponential integral function. Hence, substituting $ \Omega (\sigma _{{s_i}{d_i}}^2,\sigma _{{s_i}e}^2,{\gamma _s})$ from (31) into (26) gives
\begin{equation}\label{equa32}
\Pr (|{h_{{s_j}e}}{|^2}{\gamma _s} + 2 < 2Z) =  \frac{{2\sigma _{{s_i}e}^2}\exp (\varphi )Ei( \varphi )}{{\sigma _{{s_i}{d_i}}^2\sigma _{{s_j}e}^2{\gamma _s}}}.
\end{equation}
Finally, combining (21) and (32), we obtain the intercept probability of the SC-RJS as
\begin{equation}\label{equa33}
P_{{\mathop{\textrm{int}}} }^{{\textrm{SC-RJ}}} = \sum\limits_{i = 1}^N {\frac{{{\alpha _i}}}{{N - 1}}\sum\limits_{{S_j} \in \{ {\cal S} - {S_i}\} } {\left( \frac{{2\sigma _{{s_i}e}^2}\exp (\varphi )Ei( \varphi )}{{\sigma _{{s_i}{d_i}}^2\sigma _{{s_j}e}^2{\gamma _s}}} \right)} },
\end{equation}
where $\varphi$ is given by (30).

\subsection{SC-OJS Scheme}
In this subsection, we analyze the intercept probability of SC-OJS scheme. Similarly to (17), we obtain an intercept probability of spectrum sharing systems relying on the proposed SC-OJS scheme as
\begin{equation}\label{equa34}
P_{{\mathop{\textrm{int}}} }^{{\textrm{SC-OJ}}} = \sum\limits_{i = 1}^N {{\alpha _i}\Pr (C_{{s_i}{d_i}}^{{\textrm{SC}}} < C_{{s_i}e}^{{\textrm{SC-OJ}}})},
\end{equation}
where $C_{{s_i}{d_i}}^{{\textrm{SC}}} $ and $C_{{s_i}e}^{{\textrm{SC-OJ}}}$ are given by (7) and (13), respectively. Combining (7), (13) and (34) gives
\begin{equation}\label{equa35}
\begin{split}
P_{{\mathop{\textrm{int}}} }^{{\textrm{SC-OJ}}} &= \sum\limits_{i = 1}^N {{\alpha _i}\Pr \left[ \begin{split}
&{{\log }_2}(1 + \frac{{|{h_{{s_i}{d_i}}}{|^2}{\gamma _s}}}{2}) \\
&< \mathop {\min }\limits_{{S_j} \in \{ S - {S_i}\} } {{\log }_2}(1 + \frac{{|{h_{{s_i}e}}{|^2}{\gamma _s}}}{{|{h_{{s_j}e}}{|^2}{\gamma _s} + 2}})
\end{split}
\right]}  \\
&= \sum\limits_{i = 1}^N {{\alpha _i}\Pr \left[
\begin{split}
&{{\log }_2}(1 + \frac{{|{h_{{s_i}{d_i}}}{|^2}{\gamma _s}}}{2}) \\
&< {{\log }_2}(1 + \frac{{|{h_{{s_i}e}}{|^2}{\gamma _s}}}{{\mathop {\max }\limits_{{S_j} \in \{ S - {S_i}\} } |{h_{{s_j}e}}{|^2}{\gamma _s} + 2}})
\end{split}
\right]}  \\
&= \sum\limits_{i = 1}^N {\alpha _i}\Pr (\frac{{|{h_{{s_i}{d_i}}}{|^2}{\gamma _s}}}{2} < \frac{{|{h_{{s_i}e}}{|^2}{\gamma _s}}}{{\mathop {\max }\limits_{{S_j} \in \{ S - {S_i}\} } |{h_{{s_j}e}}{|^2}{\gamma _s} + 2}}),  \\
\end{split}
\end{equation}
which is rewritten as
\begin{equation}\label{equa36}
P_{{\mathop{\textrm{int}}} }^{{\textrm{SC-OJ}}} = \sum\limits_{i = 1}^N {{\alpha _i}\Pr (\mathop {\max }\limits_{{S_j} \in \{ {\cal S} - {S_i}\} } |{h_{{s_j}e}}{|^2}{\gamma _s} + 2 < \frac{{2|{h_{{s_i}e}}{|^2}}}{{|{h_{{s_i}{d_i}}}{|^2}}})}.
\end{equation}
Denoting $|{h_{{s_i}e}}{|^2} = X$, $|{h_{{s_i}{d_i}}}{|^2} = Y$, and $Z = \frac{X}{Y}$, we have
\begin{equation}\label{equa37}
P_{{\mathop{\textrm{int}}} }^{{\textrm{SC-OJ}}} = \sum\limits_{i = 1}^N {{\alpha _i}\Pr (\mathop {\max }\limits_{{S_j} \in \{ {\cal S} - {S_i}\} } |{h_{{s_j}e}}{|^2}{\gamma _s} + 2 < 2Z)}.
\end{equation}
Noting again that random variable $|{h_{{s_j}e}}{|^2}$ is exponentially distributed and independent of $Z$, we obtain (37) as
\begin{equation}\label{equa38}
P_{{\mathop{\textrm{int}}} }^{{\textrm{SC-OJ}}} = \sum\limits_{i = 1}^N {{\alpha _i}\int_1^\infty  {\prod\limits_{{S_j} \in \{ {\cal S} - {S_i}\} } {[1 - \exp ( - \frac{{2z - 2}}{{\sigma _{{s_j}e}^2{\gamma _s}}})]} {p_Z}(z)dz} },
\end{equation}
where $P_Z(z)$ is the PDF of random variable $Z$ as given by (25). Using the result of Appendix A, we obtain the intercept probability of SC-OJS scheme from (38) as
\begin{equation}\label{equa39}
P_{{\mathop{\textrm{int}}} }^{{\textrm{SC-OJ}}} =  \sum\limits_{i = 1}^N {{\alpha _i}[\sum\limits_{k = 1}^{{2^{N - 1}} - 1} \sum\limits_{{S_j} \in {{\cal J}_k}} {\frac{{{{( - 1)}^{{|{\cal J}_k}|+1}}2\sigma _{{s_i}e}^2}}{{\sigma _{{s_i}{d_i}}^2\sigma _{{s_j}e}^2{\gamma _s}}}} \exp (\phi )Ei(  \phi ) ]},
\end{equation}
where the parameter $\phi $ is defined as
\begin{equation}\label{equa40}
\phi  = \frac{{2\sigma _{{s_i}{d_i}}^2 + 2\sigma _{{s_i}e}^2}}{{\sigma _{{s_i}{d_i}}^2{\gamma _s}}}(\sum\limits_{{S_j} \in {{\cal J}_k}} {\frac{1}{{\sigma _{{s_j}e}^2}}} ),
\end{equation}
where ${\cal J}_k$ represents the $k$-th non-empty subcollection of the set $\{{\cal S} - {S_i}\}$. As shown in (16), (33) and (39), we have now derived closed-form intercept probability expressions for the conventional non-cooperation as well as the proposed SC-RJS and SC-OJS schemes over Rayleigh fading channels.

\section{Secrecy Diversity Gain Analysis}
In this section, we present the secrecy diversity analysis for the conventional non-cooperation, SC-RJS, and SC-OJS schemes in high SNR region. Although the closed-form intercept probability expressions as given by (16), (33) and (39) can be used for numerical performance evaluation, they fail to provide an insight into the impact of the number of {{source-destination pairs}} on the physical-layer security of spectrum sharing systems.

\subsection{Conventional Non-cooperation}
This subsection conducts an asymptotic intercept probability analysis of conventional non-cooperation scheme and presents its secrecy diversity gain as a baseline. As discussed in [26], the traditional diversity gain is introduced to measure the reliability of wireless communications, which is mathematically defined as
\begin{equation}\label{equa41}
d =  - \mathop {\lim }\limits_{\gamma_s \to \infty } \frac{{\log P_e (\gamma_s)}}{{\log {\gamma_s}}},
\end{equation}
where $\gamma_s$ represents the SNR and ${P_e (\gamma_s)}$ represents the bit error rate (BER) as a function of $\gamma_s$. From (41), one can observe that the BER behaves as $\frac{1}{{\gamma_s}^d}$ for ${{\gamma_s} \to \infty } $, implying that with an increasing diversity gain $d$, the BER is reduced faster in high SNR region. Similarly to (41), we introduce a secrecy diversity gain to characterize an asymptotic behavior of the intercept probability in high SNR, which is defined as a ratio of the logarithmic intercept probability to the logarithmic SNR $\gamma_s$, i.e.,
\begin{equation}\label{equa42}
d_s =  - \mathop {\lim }\limits_{\gamma_s \to \infty } \frac{{\log P_{\textrm{int}} (\gamma_s)}}{{\log {\gamma_s}}},
\end{equation}
where ${P_{\textrm{int}} (\gamma_s)}$ represents the intercept probability as a function of $\gamma_s$. From (42), we obtain a secrecy diversity gain of the non-cooperation scheme as
\begin{equation}\label{equa43}
d_s^{{\textrm{nonC}}} =  - \mathop {\lim }\limits_{{\gamma_s} \to \infty } \frac{{\log P_{{\mathop{\textrm{int}}} }^{{\textrm{nonC}}}}}{{\log {\gamma _s}}},
\end{equation}
where $P_{{\mathop{\textrm{int}}} }^{{\textrm{nonC}}}$ represents the intercept probability of conventional non-cooperation scheme. Substituting $P_{{\mathop{\textrm{int}}} }^{{\textrm{nonC}}}$ from (16) into (43) yields
\begin{equation}\label{equa44}
d_s^{{\textrm{nonC}}} =  - \mathop {\lim }\limits_{{\gamma _s} \to \infty } \frac{{\log (\sum\limits_{i = 1}^N {\frac{{{\alpha _i}\sigma _{{s_i}e}^2}}{{\sigma _{{s_i}{d_i}}^2 + \sigma _{{s_i}e}^2}}} )}}{{\log {\gamma _s}}} = 0,
\end{equation}
which shows that no secrecy diversity is achieved by the conventional non-cooperation. Again, this implies that increasing the transmit power $P_s$ would not improve the physical-layer security of spectrum sharing systems with the non-cooperation scheme in terms of its intercept probability.

\subsection{SC-RJS Scheme}
In this subsection, we present the secrecy diversity analysis of the SC-RJS scheme. Using (42), we obtain a secrecy diversity gain of the SC-RJS scheme as
\begin{equation}\label{equa45}
d_s^{{\textrm{SC-RJ}}} =  - \mathop {\lim }\limits_{{\gamma _s} \to \infty } \frac{{\log P_{{\mathop{\textrm{int}}} }^{{\textrm{SC-RJ}}}}}{{\log {\gamma _s}}},
\end{equation}
where ${P_{{\mathop{\textrm{int}}} }^{{\textrm{SC-RJ}}}}$ is given by (33). Following [27, Eq. 5.1.20], $Ei(\phi)$ is bounded to
\begin{equation}\label{equa46}
\frac{1}{2}\exp ( - \varphi )\ln (1 + \frac{2}{\varphi }) \le Ei(\varphi ) \le \exp ( - \varphi )\ln (1 + \frac{1}{\varphi }),
\end{equation}
for $\varphi > 0$. Combining (33) and (46), we have
\begin{equation}\label{equa47}
P_{{\mathop{\textrm{int}}} }^{{\textrm{SC-RJ}}} \le \sum\limits_{i = 1}^N {\frac{{{\alpha _i}}}{{N - 1}}\sum\limits_{{S_j} \in \{ {\cal S} - {S_i}\} } {[\frac{{2\sigma _{{s_i}e}^2\ln (1 + {\varphi ^{ - 1}})}}{{\sigma _{{s_i}{d_i}}^2\sigma _{{s_j}e}^2{\gamma _s}}}]} }.
\end{equation}
Substituting $\varphi $ from (30) into (47) yields
\begin{equation}\label{equa48}
P_{{\mathop{\textrm{int}}} }^{{\textrm{SC-RJ}}} \le \sum\limits_{i = 1}^N {\frac{{{\alpha _i}}}{{N - 1}}\sum\limits_{{S_j} \in \{ {\cal S} - {S_i}\} } {[\frac{{2\sigma _{{s_i}e}^2\ln (1 + \frac{{\sigma _{{s_i}{d_i}}^2\sigma _{{s_j}e}^2{\gamma _s}}}{{2\sigma _{{s_i}{d_i}}^2 + 2\sigma _{{s_i}e}^2}})}}{{\sigma _{{s_i}{d_i}}^2\sigma _{{s_j}e}^2{\gamma _s}}}]} }.
\end{equation}
Letting ${{\gamma _s} \to \infty }$, we rewrite (48) as
\begin{equation}\label{equa49}
\begin{split}
\mathop {\lim }\limits_{{\gamma _s} \to \infty } P_{{\mathop{\textrm{int}}} }^{{\textrm{SC-RJ}}} &\le \sum\limits_{i = 1}^N {\frac{{{\alpha _i}}}{{N - 1}}\sum\limits_{{S_j} \in \{ {\cal S} - {S_i}\} } {[\frac{{2\sigma _{{s_i}e}^2\ln ({\gamma _s})}}{{\sigma _{{s_i}{d_i}}^2\sigma _{{s_j}e}^2{\gamma _s}}}]} }  \\
&= [\sum\limits_{i = 1}^N {\frac{{{\alpha _i}}}{{N - 1}}\sum\limits_{{S_j} \in \{ {\cal S} - {S_i}\} } {(\frac{{2\sigma _{{s_i}e}^2}}{{\sigma _{{s_i}{d_i}}^2\sigma _{{s_j}e}^2}})} } ] \cdot \frac{{\ln ({\gamma _s})}}{{{\gamma _s}}}.
\end{split}
\end{equation}
Combining (45) and (49), we arrive at
\begin{equation}\label{equa50}
\begin{split}
d_s^{{\textrm{SC-RJ}}} \ge & 1 - \mathop {\lim }\limits_{{\gamma _s} \to \infty } \frac{{\log [\sum\limits_{i = 1}^N {\frac{{{\alpha _i}}}{{N - 1}}\sum\limits_{{S_j} \in \{ {\cal S} - {S_i}\} } {(\frac{{2\sigma _{{s_i}e}^2}}{{\sigma _{{s_i}{d_i}}^2\sigma _{{s_j}e}^2}})} } ]}}{{\log {\gamma _s}}}\\
&- \mathop {\lim }\limits_{{\gamma _s} \to \infty } \frac{{\log [\ln ({\gamma _s})]}}{{\log {\gamma _s}}}.
\end{split}
\end{equation}
Considering ${{\gamma _s} \to \infty }$, we have
\begin{equation}\label{equa51}
\mathop {\lim }\limits_{{\gamma _s} \to \infty } \frac{{\log [\sum\limits_{i = 1}^N {\frac{{{\alpha _i}}}{{N - 1}}\sum\limits_{{S_j} \in \{ {\cal S} - {S_i}\} } {(\frac{{2\sigma _{{s_i}e}^2}}{{\sigma _{{s_i}{d_i}}^2\sigma _{{s_j}e}^2}})} } ]}}{{\log {\gamma _s}}} = 0,
\end{equation}
and
\begin{equation}\label{equa52}
\mathop {\lim }\limits_{{\gamma _s} \to \infty } \frac{{\log [\ln ({\gamma _s})]}}{{\log {\gamma _s}}} = 0.
\end{equation}
Substituting (51) and (52) into (50) gives
\begin{equation}\label{equa53}
d_s^{{\textrm{SC-RJ}}} \ge 1.
\end{equation}
Additionally, using (33) and (46), we obtain
\begin{equation}\label{equa54}
P_{{\mathop{\textrm{int}}} }^{{\textrm{SC-RJ}}} \ge \sum\limits_{i = 1}^N {\frac{{{\alpha _i}}}{{N - 1}}\sum\limits_{{S_j} \in \{ {\cal S} - {S_i}\} } {[\frac{{\sigma _{{s_i}e}^2\ln (1 + 2{\varphi ^{ - 1}})}}{{\sigma _{{s_i}{d_i}}^2\sigma _{{s_j}e}^2{\gamma _s}}}]} }.
\end{equation}
Substituting $\varphi $ from (30) into (54) gives
\begin{equation}\label{equa55}
P_{{\mathop{\textrm{int}}} }^{{\textrm{SC-RJ}}} \ge \sum\limits_{i = 1}^N {\frac{{{\alpha _i}}}{{N - 1}}\sum\limits_{{S_j} \in \{ {\cal S} - {S_i}\} } {[\frac{{\sigma _{{s_i}e}^2\ln (1 + \frac{{2\sigma _{{s_i}{d_i}}^2\sigma _{{s_j}e}^2{\gamma _s}}}{{2\sigma _{{s_i}{d_i}}^2 + 2\sigma _{{s_i}e}^2}})}}{{\sigma _{{s_i}{d_i}}^2\sigma _{{s_j}e}^2{\gamma _s}}}]} },
\end{equation}
from which we have
\begin{equation}\label{equa56}
\begin{split}
 \mathop {\lim }\limits_{{\gamma _s} \to \infty } P_{{\mathop{\textrm{int}}} }^{{\textrm{SC-RJ}}} &\ge \sum\limits_{i = 1}^N {\frac{{{\alpha _i}}}{{N - 1}}\sum\limits_{{S_j} \in \{ {\cal S} - {S_i}\} } {[\frac{{\sigma _{{s_i}e}^2\ln ({\gamma _s})}}{{\sigma _{{s_i}{d_i}}^2\sigma _{{s_j}e}^2{\gamma _s}}}]} }  \\
&= (\sum\limits_{i = 1}^N {\frac{{{\alpha _i}}}{{N - 1}}\sum\limits_{{S_j} \in \{ {\cal S} - {S_i}\} } {\frac{{\sigma _{{s_i}e}^2}}{{\sigma _{{s_i}{d_i}}^2\sigma _{{s_j}e}^2}})} }  \cdot \frac{{\ln ({\gamma _s})}}{{{\gamma _s}}}. \\
 \end{split}
\end{equation}
Combining (45) and (56), we arrive at
\begin{equation}\label{equa57}
\begin{split}
d_s^{{\textrm{SC-RJ}}} \le& 1 - \mathop {\lim }\limits_{{\gamma _s} \to \infty } \frac{{\log (\sum\limits_{i = 1}^N {\frac{{{\alpha _i}}}{{N - 1}}\sum\limits_{{S_j} \in \{ {\cal S} - {S_i}\} } {\frac{{\sigma _{{s_i}e}^2}}{{\sigma _{{s_i}{d_i}}^2\sigma _{{s_j}e}^2}})} } }}{{\log {\gamma _s}}}\\
&- \mathop {\lim }\limits_{{\gamma _s} \to \infty } \frac{{\log [\ln ({\gamma _s})]}}{{\log {\gamma _s}}}.
\end{split}
\end{equation}
Letting ${{\gamma _s} \to \infty }$, we have
\begin{equation}\label{equa58}
\mathop {\lim }\limits_{{\gamma _s} \to \infty } \frac{{\log (\sum\limits_{i = 1}^N {\frac{{{\alpha _i}}}{{N - 1}}\sum\limits_{{S_j} \in \{ {\cal S} - {S_i}\} } {\frac{{\sigma _{{s_i}e}^2}}{{\sigma _{{s_i}{d_i}}^2\sigma _{{s_j}e}^2}})} } }}{{\log {\gamma _s}}} = 0.
\end{equation}
Substituting (52) and (58) into (57) yields
\begin{equation}\label{equa59}
d_s^{{\textrm{SC-RJ}}} \le 1.
\end{equation}
Finally, using the squeeze theorem, we obtain the secrecy diversity gain of SC-RJS scheme from (53) and (59) as
\begin{equation}\label{equa60}
d_s^{{\textrm{SC-RJ}}} = 1,
\end{equation}
which shows that the intercept probability behaves as $\frac{1}{{\gamma_s}}$ in high SNR region. This means that the intercept probability of the SC-RJS can be notably reduced with an increasing transmit power, showing its secrecy advantage over the conventional non-cooperation scheme.

\subsection{SC-OJS Scheme}
In this subsection, we analyze the secrecy diversity of the SC-OJS. Following (42), a secrecy diversity gain of the SC-OJS scheme is obtained as
\begin{equation}\label{equa61}
d_s^{{\textrm{SC-OJ}}} =  - \mathop {\lim }\limits_{{\gamma _s} \to \infty } \frac{{\log (P_{{\mathop{\textrm{int}}} }^{{\textrm{SC-OJ}}})}}{{\log {\gamma _s}}},
\end{equation}
where ${P_{{\mathop{\textrm{int}}} }^{{\textrm{SC-OJ}}}}$ is given by (39). Similarly to (46), we have
\begin{equation}\label{equa62}
\frac{1}{2}\exp ( - \phi )\ln (1 + \frac{2}{\phi }) \le Ei(\phi ) \le \exp ( - \phi )\ln (1 + \frac{1}{\phi }),
\end{equation}
where $\phi$ is given by (40). Considering ${{\gamma _s} \to \infty }$ and using (40), we obtain
\begin{equation}\label{equa63}
\mathop {\lim }\limits_{{\gamma _s} \to \infty } \exp ( - \phi ) = 1,
\end{equation}
and
\begin{equation}\label{equa64}
\mathop {\lim }\limits_{{\gamma _s} \to \infty } \ln (1 + \frac{1}{\phi }) = \ln ({\gamma _s}),
\end{equation}
and
\begin{equation}\label{equa65}
\mathop {\lim }\limits_{{\gamma _s} \to \infty } \ln (1 + \frac{2}{\phi }) = \ln ({\gamma _s}).
\end{equation}
Combining (62)-(65), we arrive at
\begin{equation}\label{equa66}
\ln ({\gamma _s}) \le \mathop {\lim }\limits_{{\gamma _s} \to \infty } Ei(\phi ) \le \ln ({\gamma _s}),
\end{equation}
which in turn leads to
\begin{equation}\label{equa67}
\mathop {\lim }\limits_{{\gamma _s} \to \infty } Ei(\phi ) = \ln ({\gamma _s}).
\end{equation}
Moreover, letting ${{\gamma _s} \to \infty }$, we similarly obtain
\begin{equation}\label{equa68}
\mathop {\lim }\limits_{{\gamma _s} \to \infty } \exp ( \phi ) = 1.
\end{equation}
Substituting (67) and (68) into (39), we have
\begin{equation}\label{equa69}
\begin{split}
\mathop {\lim }\limits_{{\gamma _s} \to \infty } P_{{\mathop{\textrm{int}}} }^{{\textrm{SC-OJ}}} = & \sum\limits_{i = 1}^N {\frac{{2{\alpha _i}\sigma _{{s_i}e}^2}}{{\sigma _{{s_i}{d_i}}^2}}[ - \sum\limits_{k = 1}^{{2^{N - 1}} - 1} {{{( - 1)}^{{\rm{|}}{{\cal J}_k}{\rm{|}}}}\sum\limits_{{S_j} \in {{\cal J}_k}} {\frac{1}{{\sigma _{{s_j}e}^2}}} } ]}  \\
&\quad\quad \times \frac{{\ln ({\gamma _s})}}{{{\gamma _s}}}.
\end{split}
\end{equation}
Combining (61) and (69) yields (70) at the top of the following page.
\begin{figure*}
\begin{equation}\label{equa70}
d_s^{{\textrm{SC-OJ}}} = 1 - \mathop {\lim }\limits_{{\gamma _s} \to \infty } \frac{{\log \left( {\sum\limits_{i = 1}^N {\frac{{2{\alpha _i}\sigma _{{s_i}e}^2}}{{\sigma _{{s_i}{d_i}}^2}}[ - \sum\limits_{k = 1}^{{2^{N - 1}} - 1} {{{( - 1)}^{{\rm{|}}{{\cal J}_k}{\rm{|}}}}\sum\limits_{{S_j} \in {{\cal J}_k}} {\frac{1}{{\sigma _{{s_j}e}^2}}} } ]} } \right)}}{{\log {\gamma _s}}} - \mathop {\lim }\limits_{{\gamma _s} \to \infty } \frac{{\log [\ln ({\gamma _s})]}}{{\log {\gamma _s}}}.
\end{equation}
\end{figure*}
Clearly, one can readily obtain
\begin{equation}\label{equa71}
\mathop {\lim }\limits_{{\gamma _s} \to \infty } \frac{{\log \left( {\sum\limits_{i = 1}^N {\frac{{2{\alpha _i}\sigma _{{s_i}e}^2}}{{\sigma _{{s_i}{d_i}}^2}}[ - \sum\limits_{k = 1}^{{2^{N - 1}} - 1} {{{( - 1)}^{{\rm{|}}{{\cal J}_k}{\rm{|}}}}\sum\limits_{{S_j} \in {{\cal J}_k}} {\frac{1}{{\sigma _{{s_j}e}^2}}} } ]} } \right)}}{{\log {\gamma _s}}} = 0.
\end{equation}
Therefore, substituting (52) and (71) into (70) gives
\begin{equation}\label{equa72}
d_s^{{\textrm{SC-OJ}}} = 1,
\end{equation}
which shows that the secrecy diversity gain of one is achieved by the SC-OJS scheme. One can observe from (60) and (72) that the SC-RJS and SC-OJS schemes achieve the same secrecy diversity gain. This surprisingly means that the optimal jammer selection fails to provide a further performance improvement compared to the random jammer selection in terms of the secrecy diversity gain.

\section{Numerical Results and Discussions}
This section presents numerical intercept probability results of the conventional non-cooperation as well as the proposed SC-RJS and SC-OJS schemes by using (16), (33) and (39). In our numerical evaluation, the duty cycle of $\alpha_i=1/N$ is considered for different {{source-destination pairs}} and the average gains are specified to $\sigma^2_{s_id_i}=\sigma^2_{s_ie}=\sigma^2_{s_je}=1$, unless otherwise stated. For notational convenience, let $\lambda=\sigma^2_{s_id_i}/\sigma^2_{s_ie}$ denote the ratio of the average gains between the main channel and eavesdropping channel, referred to as the main-to-eavesdropping ratio (MER). Additionally, the number of {{source-destination pairs}} $N=4$ is used, unless otherwise mentioned.

\begin{figure}
  \centering
  {\includegraphics[scale=0.55]{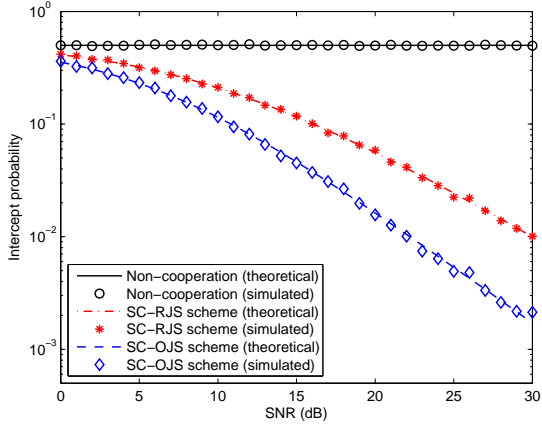}\\
  \caption{Intercept probability versus SNR $\gamma_s$ of the conventional non-cooperation as well as the proposed SC-RJS and SC-OJS schemes.}\label{Fig2}}
\end{figure}
Fig. 2 shows the intercept probability comparison among the conventional non-cooperation, the SC-RJS, and the SC-OJS schemes by plotting (16), (33) and (39) as a function of the SNR $\gamma_s$. The simulated intercept probability results are also given in Fig. 2, where the continuous lines and discrete markers are used to represent the theoretical and simulated intercept probability results, respectively. It can be seen from Fig. 2 that as the SNR $\gamma_s$ increases, the intercept probability of conventional non-cooperation scheme keeps unchanged, as implied from (16). By contrast, with an increasing SNR, the intercept probabilities of proposed SC-RJS and SC-OJS schemes are reduced significantly. This shows the physical-layer security benefits of exploiting the source cooperation against eavesdropping, as compared to the conventional non-cooperation. Additionally, one can observe from Fig. 2 that the theoretical intercept probabilities of the non-cooperation, SC-RJS and SC-OJS schemes match well with the corresponding simulation results, confirming the correctness of our closed-form intercept probability expressions of (16), (33) and (39).

\begin{figure}
  \centering
  {\includegraphics[scale=0.55]{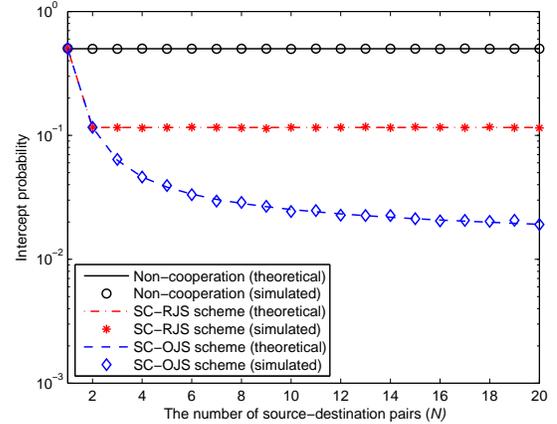}\\
  \caption{{Intercept probability versus the number of {{source-destination pairs}} $N$ of the conventional non-cooperation as well as the proposed SC-RJS and SC-OJS schemes.}}\label{Fig2}}
\end{figure}
Fig. 3 depicts the intercept probability versus the number of {{source-destination pairs}} $N$ of the conventional non-cooperation as well as the proposed SC-RJS and SC-OJS schemes. As shown in Fig. 3, both the theoretical and simulated intercept probability results match each other, which further validates our closed-form intercept probability analysis. One can also see from Fig. 3 that with an increasing number of source-destination pairs, the intercept probability performance of the conventional non-cooperation remains the same, whereas the intercept probability of the SC-RJS decreases when $N$ increases from $N=1$ to $2$ and then becomes stable as the number of {{source-destination pairs}} $N$ continues to increase thereafter. This is because that given $N=1$ (i.e. there is only one {{source-destination pair}}), the source cooperation is unavailable and thus the intercept performance of the SC-RJS in this case becomes identical to that of the conventional non-cooperation. When $N$ increases from $N=1$ to $2$, it becomes available to exploit the SC strategy for decreasing the intercept probability. Moreover, as the number of {{source-destination pairs}} continues to increase more than two, a randomly selected source node is allowed in the RJS to act as a friendly jammer, which is not beneficial to the physical-layer security improvement. By contrast, the OJS scheme allows an optimal source node to be chosen as the friendly jammer for minimizing the confidential information leakage, hence the intercept probability of the SC-OJS always decreases with an increasing number of {{source-destination pairs}}, as can be observed from Fig. 3.

\begin{figure}
  \centering
  {\includegraphics[scale=0.55]{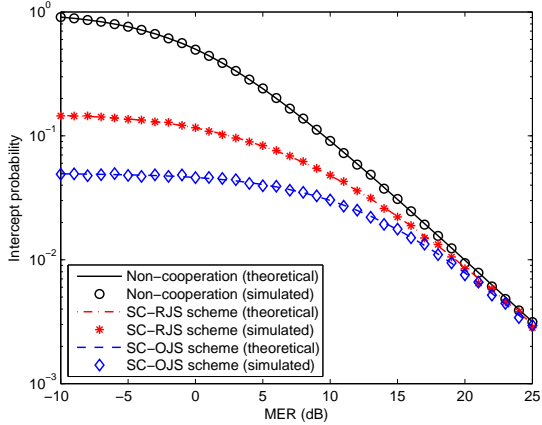}\\
  \caption{Intercept probability versus MER $\lambda$ of the conventional non-cooperation as well as the proposed SC-RJS and SC-OJS schemes.}\label{Fig4}}
\end{figure}

Fig. 4 shows the intercept probability versus MER $\lambda$ of the conventional non-cooperation as well as the proposed SC-RJS and SC-OJS schemes. It can be seen from Fig. 4 that as the MER increases, the intercept performance of the non-cooperation, SC-RJS and SC-OJS improves accordingly, which is because that the eavesdropping channel worsens with an increasing MER $\lambda$. One can also observe from Fig. 4 that in the low MER region, the proposed SC-RJS and SC-OJS significantly outperform the conventional non-cooperation in terms of intercept probability. Moreover, as the MER increases, the intercept probabilities of the conventional non-cooperation as well as the proposed SC-RJS and SC-OJS schemes converge to each other. This is due to the fact that in the high MER region, the eavesdropping channel is much worse than the main channel and the jamming signal received at the eavesdropper may become negligible compared to the background noise, thus the security benefit of exploiting SC in high MER region is marginal.

\begin{figure}
  \centering
  {\includegraphics[scale=0.55]{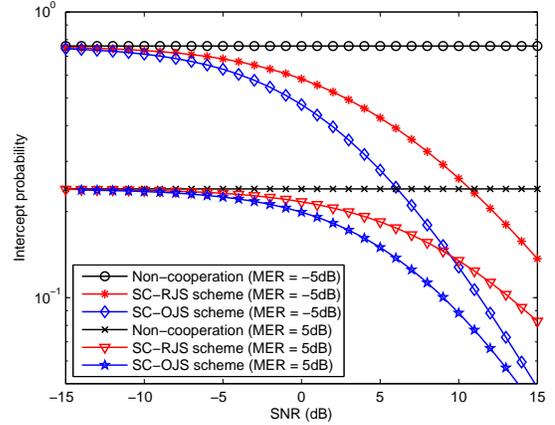}\\
  \caption{Intercept probability versus SNR of the conventional non-cooperation as well as the proposed SC-RJS and SC-OJS schemes for different MER $\lambda$.}\label{Fig5}}
\end{figure}
In Fig. 5, we demonstrate the intercept probability versus SNR of the conventional non-cooperation as well as the proposed SC-RJS and SC-OJS schemes for different MER $\lambda$. As shown in Fig. 5, for both the cases of ${\textrm{MER}}=-5{\textrm{dB}}$ and $5{\textrm{dB}}$, the conventional non-cooperation performs the worst and the proposed SC-OJS scheme is the best in terms of intercept probability. It can also be observed from Fig. 5 that with an increasing SNR, the intercept probability of the conventional non-cooperation remains constant, while the intercept performance of the SC-RJS and SC-OJS improves significantly. This means that even when the eavesdropping channel is better than the main channel (e.g., ${\textrm{MER}}=-5{\textrm{dB}}$), the physical-layer security of spectrum sharing systems relying on the SC-RJS and SC-OJS schemes can be enhanced by simply increasing the transmit power.

\begin{figure}
  \centering
  {\includegraphics[scale=0.55]{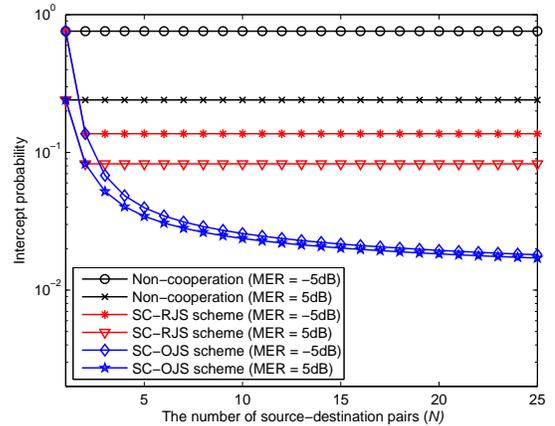}\\
  \caption{Intercept probability versus the number of {{source-destination pairs}} $N$ of the conventional non-cooperation as well as the proposed SC-RJS and SC-OJS schemes for different MER $\lambda$.}\label{Fig6}}
\end{figure}
Fig. 6 shows the intercept probability versus the number of {{source-destination pairs}} $N$ of the conventional non-cooperation as well as the proposed SC-RJS and SC-OJS schemes for different MER $\lambda$. One can observe from Fig. 6 that for both the cases of ${\textrm{MER}}=-5{\textrm{dB}}$ and $5{\textrm{dB}}$, the intercept probabilities of the non-cooperation and SC-RJS are independent of the number of {{ source-destination pairs}} $N$, whereas the intercept performance of the SC-OJS is slightly improved with an increasing $N$. Therefore, increasing the number of {{source-destination pairs}} is beneficial to the physical-layer security of the SC-OJS, even if the main channel is much worse than the eavesdropping channel (e.g., ${\textrm{MER}}=-5{\textrm{dB}}$). However, the secrecy enhancement of the SC-OJS by increasing the number of {{source-destination pairs}} is incremental, as seen from Fig. 6.

\section{Conclusions}
In this paper, we have investigated the physical-layer security for a spectrum sharing system consisting of {{multiple source-destination pairs}}, each consisting of a source node transmitting to its destination, where an eavesdropper is considered to tap an active transmission between any source-destination pairs. We have explored a source cooperation (SC) aided opportunistic jamming framework for protecting the spectrum sharing system against eavesdropping. More specifically, when a {{source node}} is allowed to access the shared spectrum for data transmissions, another source node is opportunistically selected to act as a friendly jammer for confusing the eavesdropper without affecting the legitimate transmissions. We have presented two SC aided opportunistic jamming methods, namely the SC-RJS and SC-OJS, and derived their intercept probability expressions in closed-form over Rayleigh fading channels. For comparison purposes, we have also considered the conventional non-cooperation as a baseline. We have carried out an asymptotic intercept probability analysis for the non-cooperation, SC-RJS and SC-OJS in the high SNR region. It has been shown that the conventional non-cooperation achieves a secrecy diversity of zero only, whereas a higher secrecy diversity of one is achieved by both the SC-RJS and SC-OJS schemes. Numerical results have demonstrated that the proposed SC-OJS performs the best and the conventional non-cooperation achieves the worst secrecy performance in terms of intercept probability.

{{It needs to be pointed out that in this paper, we have investigated a simple case where only single source-destination pair is actively transmitting at a time with the aid of a single friendly jammer in the presence of a single eavesdropper. It is of interest to explore a more general case with multiple concurrent source-destination transmissions, multiple jammers and multiple eavesdroppers. In contrast to an eavesdropper, multiple eavesdroppers can perform independently or collaboratively in tapping the legitimate transmissions, leading to an increasing intercept probability. We leave this interesting problem for future work.}}

\appendices
\section{Calculation of (39)}
By using the binomial expansion theorem, the term $\prod\limits_{{S_j} \in \{ {\cal S} - {S_i}\} } {[1 - \exp ( - \frac{{2z - 2}}{{\sigma _{{s_j}e}^2{\gamma _s}}})]}$ can be expanded as
\begin{equation}
\begin{split}
&\prod\limits_{{S_j} \in \{ {\cal S} - {S_i}\} } {[1 - \exp ( - \frac{{2z - 2}}{{\sigma _{{s_j}e}^2{\gamma _s}}})]}  \\
&= 1 + \sum\limits_{k = 1}^{{2^{N - 1}} - 1} {{{( - 1)}^{|{{\cal J}_k}|}}\exp ( - \sum\limits_{{S_j} \in {{\cal J}_k}} {\frac{{2z - 2}}{{\sigma _{{s_j}e}^2{\gamma _s}}}} )},\\
\end{split}\tag{A.1}\label{A.1}
\end{equation}
where ${\cal J}_k$ represents the $k$-th non-empty subcollection of the set $\{{\cal S} - {S_i}\}$. Combining (A.1) and (38), we arrive at
\begin{equation}
\begin{split}
P_{{\mathop{\textrm{int}}} }^{{\textrm{SC-OJ}}} = &\sum\limits_{i = 1}^N {{\alpha _i}\int_1^\infty  {[1 + \sum\limits_{k = 1}^{{2^{N - 1}} - 1} {{{( - 1)}^{|{{\cal J}_k}|}}}} } \\
&\quad \times \exp ( - \sum\limits_{{S_j} \in {{\cal J}_k}} {\frac{{2z - 2}}{{\sigma _{{s_j}e}^2{\gamma _s}}}} )]{p_Z}(z)dz,
\end{split}\tag{A.2}\label{A.2}
\end{equation}
where $p_Z(z)$ represents the PDF of $Z$. Substituting $p_Z(z)$ from (25) into (A.2) gives
\begin{equation}
\begin{split}
P_{{\mathop{\textrm{int}}} }^{{\textrm{SC-OJ}}} = &\sum\limits_{i = 1}^N {{\alpha _i}[{\Phi _1}(\sigma _{{s_i}{d_i}}^2,\sigma _{{s_i}e}^2)}\\
&\quad+ \sum\limits_{k = 1}^{{2^{N - 1}} - 1} {{{( - 1)}^{|{{\cal J}_k}|}}{\Phi _k}(\sigma _{{s_i}{d_i}}^2,\sigma _{{s_j}e}^2,\sigma _{{s_i}e}^2)} ] ,\\
\end{split}\tag{A.3}\label{A.3}
\end{equation}
where ${\Phi _1}(\sigma _{{s_i}{d_i}}^2,\sigma _{{s_i}e}^2)$ and ${\Phi _k}(\sigma _{{s_i}{d_i}}^2,\sigma _{{s_j}e}^2,\sigma _{{s_i}e}^2) $ are defined as
\begin{equation}
{\Phi _1}(\sigma _{{s_i}{d_i}}^2,\sigma _{{s_i}e}^2) = \int_1^\infty  {\frac{{\sigma _{{s_i}{d_i}}^2\sigma _{{s_i}e}^2}}{{{{(\sigma _{{s_i}{d_i}}^2z + \sigma _{{s_i}e}^2)}^2}}}dz}, \tag{A.4}\label{A.4}
\end{equation}
and
\begin{equation}
\begin{split}
{\Phi _k}(\sigma _{{s_i}{d_i}}^2,\sigma _{{s_j}e}^2,\sigma _{{s_i}e}^2) =& \int_1^\infty  {\frac{{\sigma _{{s_i}{d_i}}^2\sigma _{{s_i}e}^2}}{{{{(\sigma _{{s_i}{d_i}}^2z + \sigma _{{s_i}e}^2)}^2}}}}\\
&\times\exp ( - \sum\limits_{{S_j} \in {{\cal J}_k}} {\frac{{2z - 2}}{{\sigma _{{s_j}e}^2{\gamma _s}}}} )dz.\\
\end{split} \tag{A.5}\label{A.5}
\end{equation}
From (A.4), we can readily obtain
\begin{equation}
{\Phi _1}(\sigma _{{s_i}{d_i}}^2,\sigma _{{s_i}e}^2) = \frac{{\sigma _{{s_i}e}^2}}{{\sigma _{{s_i}{d_i}}^2 + \sigma _{{s_i}e}^2}}.\tag{A.6}\label{A.6}
\end{equation}
Additionally, letting $\sum\limits_{{S_j} \in {{\cal J}_k}} {\frac{{2z}}{{\sigma _{{s_j}e}^2{\gamma _s}}}}  + \sum\limits_{{S_j} \in {{\cal J}_k}} {\frac{{2\sigma _{{s_i}e}^2}}{{\sigma _{{s_i}{d_i}}^2\sigma _{{s_j}e}^2{\gamma _s}}}}  =  t$, we have
\begin{equation}
z =  t{(\sum\limits_{{S_j} \in {{\cal J}_k}} {\frac{2}{{\sigma _{{s_j}e}^2{\gamma _s}}}} )^{ - 1}} - \frac{{\sigma _{{s_i}e}^2}}{{\sigma _{{s_i}{d_i}}^2}}.\tag{A.7}\label{A.7}
\end{equation}
Combining (A.5) and (A.7), we can obtain
\begin{equation}
{\Phi _k}(\sigma _{{s_i}{d_i}}^2,\sigma _{{s_j}e}^2,\sigma _{{s_i}e}^2) = \sum\limits_{{S_j} \in {{\cal J}_k}} {\frac{{2\sigma _{{s_i}e}^2\exp (\phi )}}{{\sigma _{{s_i}{d_i}}^2\sigma _{{s_j}e}^2{\gamma _s}}}} \int_{ \phi }^{ \infty } {\frac{{\exp (-t)}}{{{t^2}}}dt},\tag{A.8}\label{A.8}
\end{equation}
where the parameter $\phi$ is given by
\begin{equation}
\phi  = \frac{{2\sigma _{{s_i}{d_i}}^2 + 2\sigma _{{s_i}e}^2}}{{\sigma _{{s_i}{d_i}}^2{\gamma _s}}}(\sum\limits_{{S_j} \in {{\cal J}_k}} {\frac{1}{{\sigma _{{s_j}e}^2}}} ).\tag{A.9}\label{A.9}
\end{equation}
By performing the partial integration to (A.8), the term ${\Phi _k}(\sigma _{{s_i}{d_i}}^2,\sigma _{{s_j}e}^2,\sigma _{{s_i}e}^2)$ is obtained as
\begin{equation}
\begin{split}
{\Phi _k}(\sigma _{{s_i}{d_i}}^2,\sigma _{{s_j}e}^2,\sigma _{{s_i}e}^2) =& \frac{{\sigma _{{s_i}e}^2}}{{\sigma _{{s_i}{d_i}}^2+ \sigma _{{s_i}e}^2}} \\
& - \sum\limits_{{S_j} \in {{\cal J}_k}} {\frac{{2\sigma _{{s_i}e}^2}}{{\sigma _{{s_i}{d_i}}^2\sigma _{{s_j}e}^2{\gamma _s}}}} \exp (\phi )Ei(  \phi ). \\
\end{split} \tag{A.10}\label{A.10}
\end{equation}
Finally, substituting ${\Phi _1}(\sigma _{{s_i}{d_i}}^2,\sigma _{{s_i}e}^2)$ and ${\Phi _k}(\sigma _{{s_i}{d_i}}^2,\sigma _{{s_j}e}^2,\sigma _{{s_i}e}^2)$ from (A.6) and (A.10) into (A.3) yields (A.11) at the top of the following page,
\begin{figure*}
\begin{equation}
\begin{split}
P_{{\mathop{\textrm{int}}} }^{{\textrm{SC-OJ}}} = \sum\limits_{i = 1}^N {{\alpha _i}[\frac{{\sigma _{{s_i}e}^2}}{{\sigma _{{s_i}{d_i}}^2 + \sigma _{{s_i}e}^2}} + \sum\limits_{k = 1}^{{2^{N - 1}} - 1} {{{( - 1)}^{{|{\cal J}_k}|}}\frac{{\sigma _{{s_i}e}^2}}{{\sigma _{{s_i}{d_i}}^2 + \sigma _{{s_i}e}^2}}} ]}  + \sum\limits_{i = 1}^N {{\alpha _i}[-\sum\limits_{k = 1}^{{2^{N - 1}} - 1} {{{( - 1)}^{{|{\cal J}_k}|}}\sum\limits_{{S_j} \in {{\cal J}_k}} {\frac{{2\sigma _{{s_i}e}^2}}{{\sigma _{{s_i}{d_i}}^2\sigma _{{s_j}e}^2{\gamma _s}}}} \exp (\phi )Ei( \phi )} ]}, \\
\end{split}\tag{A.11}\label{A.11}
\end{equation}
\end{figure*}
which can be further obtained as
\begin{equation}
\begin{split}
P_{{\mathop{\textrm{int}}} }^{{\textrm{SC-OJ}}} =  \sum\limits_{i = 1}^N {{\alpha _i}[\sum\limits_{k = 1}^{{2^{N - 1}} - 1} {{{( - 1)}^{{|{\cal J}_k}|+1}}\sum\limits_{{S_j} \in {{\cal J}_k}} {\frac{{2\sigma _{{s_i}e}^2\exp (\phi )Ei(  \phi )}}{{\sigma _{{s_i}{d_i}}^2\sigma _{{s_j}e}^2{\gamma _s}}}} } ]},
 \end{split}\tag{A.12}\label{A.12}
\end{equation}
which completes the proof of (39).

\begin{IEEEbiography}[{\includegraphics[width=1in,height=1.25in]{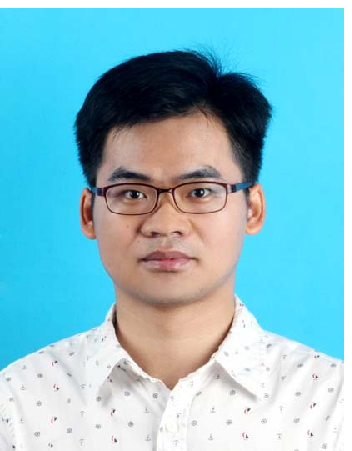}}]{Yulong Zou} (SM'13) is a Professor at the Nanjing University of Posts and Telecommunications (NUPT), Nanjing, China. He received the B.Eng. degree in information engineering from NUPT, Nanjing, China, in July 2006, the first Ph.D. degree in electrical engineering from the Stevens Institute of Technology, New Jersey, USA, in May 2012, and the second Ph.D. degree in signal and information processing from NUPT, Nanjing, China, in July 2012.

His research interests span a wide range of topics in wireless communications and signal processing, including the cooperative communications, cognitive radio, wireless security, and energy-efficient communications. Dr. Zou was awarded the 9th IEEE Communications Society Asia-Pacific Best Young Researcher in 2014. He has served as an editor for the IEEE Communications Surveys \& Tutorials, IEEE Communications Letters, IET Communications, and China Communications. In addition, he has acted as TPC members for various IEEE sponsored conferences, e.g., IEEE ICC/GLOBECOM/WCNC/VTC/ICCC, etc.

\end{IEEEbiography}

\end{document}